\documentclass[11pt]{article}

\usepackage{amssymb}
\usepackage{citesort}
\usepackage{amsmath,mathrsfs}
\usepackage{url}

\def\pa{\partial}
\def\fr{\frac}
\def\th{\theta}
\def\al{\alpha}

\def\ii{\textrm i}
\def\ee{\textrm e}

\begin{document}

\vspace*{1.0cm}
\noindent
{\bf
{\large
\begin{center}
Gauge invariant accounts of the Higgs mechanism
\end{center}
}
}

\vspace*{.5cm}
\begin{center}
Ward Struyve{\footnote{Postdoctoral Fellow FWO.}}\\
Institute of Theoretical Physics, K.U.Leuven,\\
Celestijnenlaan 200D, B--3001 Leuven, Belgium.{\footnote{Corresponding address.}}\\
Institute of Philosophy, K.U.Leuven,\\
Kardinaal Mercierplein 2, B--3000 Leuven, Belgium.\\
E--mail: Ward.Struyve@fys.kuleuven.be.
\end{center}

\begin{abstract}
\noindent
The Higgs mechanism gives mass to Yang-Mills gauge bosons. According to the conventional wisdom, this happens through the spontaneous breaking of gauge symmetry.  Yet, gauge symmetries merely reflect a redundancy in the state description and therefore the spontaneous breaking can not be an essential ingredient. Indeed, as already shown by Higgs and Kibble, the mechanism can be explained in terms of gauge invariant variables, without invoking spontaneous symmetry breaking. In this paper, we present a general discussion of such gauge invariant treatments for the case of the Abelian Higgs model, in the context of classical field theory. We thereby distinguish between two different notions of gauge: one that takes all local transformations to be gauge and one that relates gauge to a failure of determinism.   
\end{abstract}

\renewcommand{\baselinestretch}{1.1}
\bibliographystyle{unsrt}
\bibliographystyle{plain}

\section{Introduction} 
The Higgs mechanism{\footnote{This terminology is actually unfair, since the mechanism was discovered independently by Englert and Brout \cite{englert64}, Higgs \cite{higgs64}, and by Guralnik, Hagen and Kibble \cite{guralnik64}. (For some historical accounts, see \cite{brout99,higgs07,guralnik09}.)}} provides a means for Yang-Mills gauge bosons to acquire mass. According to the usual account, this happens through the spontaneous breaking of gauge symmetry. In the context of classical field theory (to which we restrict ourselves throughout the paper), the spontaneous breaking has to do with the fact that perturbations are considered around a particular ground state which is non-invariant under the gauge symmetry. The symmetry breaking yields massless bosons, called Nambu-Goldstone bosons, which combine with the gauge bosons to form massive vector bosons. 

However, a gauge symmetry is an unphysical symmetry which merely relates different representations of the same physical state or history. It could in principle be eliminated by passing to a reformulation in terms of gauge invariant degrees of freedom. As such, the physical content of the Higgs mechanism, like the masses of the fields, must be obtainable in a manifestly gauge invariant way, without ever invoking the spontaneous breaking of gauge symmetry. 

Indeed, while Higgs' original presentation involved spontaneous symmetry breaking \cite{higgs64}, he also presented a manifestly gauge invariant treatment in a subsequent paper \cite{higgs66}. He did this for the case of the Abelian Higgs model, which concerns a local $U(1)$ symmetry. He performed a field transformation that separated gauge independent degrees of freedom from gauge degrees of freedom. By keeping only the gauge invariant variables, the local $U(1)$ was no longer present and the ground state became unique. The usual results were obtained, but without spontaneous symmetry breaking.

In extending the Higgs mechanism to non-Abelian Yang-Mills theories, Kibble employed a generalization of this field transformation \cite{kibble67}. This transformation did not isolate all the gauge variables, but just those associated to the degeneracy of the ground state. By dismissing those variables, the subgroup of gauge transformations that would otherwise be broken was factored out. This made the degeneracy of the ground state disappear. The upshot was that some vector potentials gained mass, namely those corresponding to the ``broken generators'', while the others remained massless. Weinberg later introduced the unitary gauge as a short-cut to obtain Kibble's results \cite{weinberg73}. This gauge is nowadays widely adopted for the Higgs mechanism.

While the usual account has its heuristic value, the gauge invariant treatments by Higgs and Kibble have the conceptual advantage that they do not involve the spontaneous breaking of gauge symmetry. Therefore it is rather unfortunate that they seem to be largely forgotten. They seem to be absent from most reviews and text books (one exception being Rubakov's \cite{rubakov02}, which contains a discussion of the Abelian Higgs model). They also seem to have gone unnoticed by a number of authors who recently considered similar approaches \cite{chernodub08,faddeev09,masson10,ilderton10}, as well as by the philosophy of physics community, where there is an ongoing debate concerning the meaning of spontaneous symmetry breaking of gauge symmetries and the gauge invariant content of the Higgs mechanism \cite{earman03b,earman04a,earman04b,smeenk06,healey07,lyre08}. In particular, Earman, who started this debate \cite{earman03b,earman04a,earman04b}, suggested to consider a reformulation in terms of gauge invariant variables to deal with these questions. 

In this paper, we present a general discussion of such manifestly gauge invariant treatments, for the case of the Abelian Higgs model, in the context of classical field theory.{\footnote{See \cite{perez08} for a recent discussion on the meaning of spontaneous symmetry breaking in the context of quantum field theory.}} A considerable part of the discussion will revolve around the notion of gauge symmetry. Various notions exist (see for example \cite{sundermeyer82}). All of these of course have in common that gauge equivalent states or histories are observationally indistinguishable. But observationally indistinguishable states or histories could still be regarded as physically distinct. The conceptual implications for spontaneous symmetry breaking and the Higgs mechanism of course depend on the actual notion of gauge that is employed. 

We will restrict our attention to two different notions of gauge that are customary. According to the first notion, which will be considered in the first part of the paper, both the global and local symmetries are considered to be the gauge symmetries. This notion was employed by Higgs (at least for local symmetries) when presenting his gauge invariant account. We will review this account in detail. We will also consider the spontaneous symmetry breaking of a global $U(1)$ symmetry (for a self interacting scalar field) and show that a gauge invariant account can be given in a similar way.

The second notion of gauge, which will be considered in the second half of the paper, relates gauge to a failure of determinism. This notion is customary within the context of constrained dynamics, which deals with the Hamiltonian formulation of singular Lagrangians (and which is particularly relevant for canonical quantization). This is actually the notion that Earman had in mind when he called for a reformulation in terms of gauge invariant degrees of freedom. 

According to this notion, the actual gauge group depends very much on the boundary conditions of the fields. For a natural choice of boundary conditions, we will find that not all local $U(1)$ transformations are gauge symmetries and that modding them out yields a residual symmetry isomorphic to the group $U(1)$. This entails a slight modification of Higgs' treatment. We will also consider an alternative choice of gauge invariant variables, which is better suited for this case. In particular, we illustrate this in the Hamiltonian picture, where there exists a canonical way to do this in terms of the reduced phase formalism (which was explicitly asked for by Earman \cite{earman03b,earman04a,earman04b}).

The paper is organized as follows. In section \ref{symmetryandgauge}, we first introduce some terminology, in particular that of a global and local symmetry, and of spontaneous symmetry breaking. In sections \ref{ssbgs} and  \ref{ssbls}, we respectively consider the spontaneous symmetry breaking of a global $U(1)$ and local $U(1)$ symmetry and discuss the gauge invariant treatments. In section \ref{gaugefixing}, we discuss the close relation between gauge fixed variables and gauge invariant ones, which was observed before in \cite{ilderton10,salisbury09}. In particular, we illustrate this for the unitary gauge.  In sections \ref{gaugesymindet} and \ref{hamiltonianpicture}, we turn to the alternative notion of gauge which relates gauge to indeterminism. We consider the role of the boundary conditions in establishing the gauge group. For a natural choice of boundary conditions, we then discuss the elimination of gauge freedom and how the usual results of the Higgs mechanism are obtained.

\section{Symmetry and spontaneous symmetry breaking}\label{symmetryandgauge}
\subsection{Global, local and gauge symmetries}
Let us first introduce some terminology (see \cite{earman02b,earman03a,earman04b} for details). We are concerned with field theories on Minkowski space-time for which the equations of motion are derivable from an action principle, with action 
\begin{equation}
S = \int d^4x {\mathcal L}(x,\varphi(x),{\partial_\mu \varphi}(x))\,, 
\label{t1}
\end{equation}
where ${\mathcal L}$ is the Lagrangian density, which depends on the fields $\varphi=(\varphi_1,\dots,\varphi_m)$, which we assume to be smooth, and their space-time derivatives. As such the equations of motion are given by the Euler-Lagrange equations
\begin{equation}
\partial_\mu \frac{\partial {\mathcal L} }{\partial (\partial_\mu \varphi_l) } - \frac{\partial {\mathcal L} }{\partial \varphi_l } = 0\,, \qquad l=1,\dots,m\,.
\label{t2}
\end{equation}

A {\em symmetry of the equations of motion} is considered to be a transformation that maps solutions of the equations of motion to solutions. In particular, a {\em variational symmetry}, which is a transformation of the fields and the independent variables $x$ that leaves the action invariant, will be a symmetry of the equations of motion. A finite dimensional group of variational symmetries, parametrized by $s$ real variables, is called a {\em global symmetry group}.  An infinite-dimensional group, parametrized by $s$ real functions of space-time, is called a {\em local symmetry group}. 

An example of a theory with a global symmetry group is that of a complex scalar field with mass $m$, described by the Lagrangian density 
\begin{equation}
{\mathcal L}_1 =  \pa_{\mu} \phi^* \pa^{\mu} \phi    - m^2 \phi^*\phi \,.
\label{t3}
\end{equation}
The corresponding equation of motion is $\pa_{\mu} \pa^{\mu} \phi + m^2 \phi = 0$. There is a global symmetry group, given by the global $U(1)$ transformations
\begin{equation}
\phi(x) \to {\textrm e}^{{\textrm i}\alpha}\phi(x)\,, 
\label{t4}
\end{equation}
where $\alpha$ is constant and real.

An example of a theory with a local symmetry group is that of scalar electrodynamics, which is described by the Lagrangian density
\begin{equation}
{\mathcal L}_2  = (D_{\mu} \phi)^* D^{\mu} \phi  - m^2 \phi^*\phi   - \frac{1}{4} F^{\mu \nu}F_{\mu \nu} \,,
\label{t5}
\end{equation}
where $D_{\mu}= \partial_{\mu} + \ii e A_{\mu}$ is the covariant derivative with electromagnetic vector potential $A^\mu=(A_0,A_i)$, $e$ is the charge of the scalar field, and $F_{\mu \nu} = \pa_\mu A_\nu - \pa_\nu A_\mu$ is the electromagnetic field tensor.{\footnote{Throughout the paper space-time components are denoted by Greek indices ($\mu, \nu,\dots$) and spatial components by Latin indices ($i,j,\dots$).}} The corresponding Euler-Lagrange equations are
\begin{equation}
D_{\mu}D^{\mu}\phi +  m^2 \phi= 0,\quad \partial_{\mu} F^{\mu \nu}= j^{\nu} = \ii e\left(\phi^*D^{\nu}\phi - \phi D^{\nu*} \phi^* \right)\,,
\label{t5.1}
\end{equation}
where $j^{\mu}$ is the charge current. In this case, there is a local symmetry group, given by the local $U(1)$ transformations
\begin{equation}
\phi(x) \to {\textrm e}^{\ii \alpha(x)} \phi(x) \,, \quad  A_{\mu}(x) \to   A_\mu(x) -  \frac{1}{e}\partial_\mu  \alpha(x)\,, 
\label{t6}
\end{equation}
where $\alpha(x)$ may now be space-time dependent. Writing $g=\ee^{\ii \alpha}$, this becomes
\begin{equation}
\phi(x) \to g(x) \phi(x) \,, \quad  A_{\mu}(x) \to   A_\mu(x) -  \frac{\ii}{e} g(x)\partial_\mu  g(x)^{-1}\,. 
\label{t7}
\end{equation}
Note that the $g$ should be smooth in order to preserve the smoothness of the fields. 

In the above examples, both the global and local symmetries can be regarded as gauge symmetries. They both connect observationally indistinguishable solutions (at least in a hypothetical world, where those classical field equations would hold). But while observational indistinguishability is necessary to label a symmetry as gauge, it is not sufficient. One could still regard observationally indistinguishable states or solutions as physically distinct (see \cite{healey07} for a detailed discussion of such matters). This is the case, for the alternative notion of gauge symmetry which relates gauge symmetry to a failure of determinism, and which will be discussed in detail in sections \ref{gaugesymindet} and \ref{hamiltonianpicture}. In the following sections, we will regard the global and local symmetries as gauge symmetries.

To give an idea of the differences between these notions, we can consider the above examples. In the case of the global $U(1)$ symmetry there is determinism. The initial data, which are the fields and their derivatives at some initial time, uniquely determine a solution of the equations of motion. So according to the alternative definition, there is no gauge symmetry in this case. In the case of a local $U(1)$ symmetry, determinism fails, as one can have two solutions that differ by a local $U(1)$ symmetry with $\alpha$ non-zero only from some time onwards. Hence, in this case there is gauge symmetry. As we will see in section \ref{gaugesymindet}, the gauge group will depend on the boundary conditions of the fields.

This difference between global and local symmetries is also related to Noether's theorems. According to Noether's first theorem, the existence of a global symmetry group implies the existence of $s$ conserved currents. In our example of a global $U(1)$ symmetry, the conserved current is the charge current. In the case of a local symmetry group, Noether's second theorem applies, which states that the Euler-Lagrange equations are underdetermined. That is, they are not all independent but admit $s$ relations among them. This implies indeterminism.

\subsection{Spontaneous symmetry breaking}\label{ssbhm}
Given equations of motion that exhibit a certain symmetry, particular solutions may or may not be themselves symmetric under the symmetry. In particular, the ground state (the state of lowest energy) may be degenerate, with different ground states being connected by the symmetry. When considering small perturbations around a particular ground state, the equations of motion will not possess the symmetry of the fundamental equations of motion and one speaks of spontaneous symmetry breaking.{\footnote{Note that this is definitely not the only existing characterization of spontaneous symmetry breaking. For example, Strocchi \cite{strocchi08} has more restrictive characterization, which entails in particular that the system should be infinite dimensional.}}

The spontaneous symmetry breaking of a gauge symmetry is often characterized in the same way. However, such a characterization can then merely apply on the representational level, since different ground states that are connected by a gauge symmetry are merely representations of the same physical state.

\section{Spontaneous symmetry breaking of a global $U(1)$ symmetry}\label{ssbgs}
\subsection{Usual account}\label{ssbgssa}
Consider again a scalar field $\phi$, whose Lagrangian density is now given by
\begin{equation}
{\mathcal L}_3 =  \pa_{\mu} \phi^* \pa^{\mu} \phi    - V(\phi^*\phi)\,, 
\label{s1}
\end{equation}
with
\begin{equation}
V= \lambda \left(\phi^*\phi - \frac{v^2}{2}  \right)^2  =   - \mu^2\phi^*\phi + \lambda (\phi^*\phi)^2 + \frac{\mu^4}{4\lambda}   
\label{s1.1}
\end{equation}
the Higgs potential, where $\lambda$ and $v=\mu/\sqrt{\lambda}$ are both real and positive. The mass squared is negative ($m^2 = -\mu^2 < 0$), so that the field is tachyonic. The constant $\mu^4/4\lambda$ is introduced to ensure that the states of interest (such as the ground states) have a finite action and energy. 

There is a global symmetry group of transformations  $\phi(x)  \to {\textrm e}^{{\textrm i}\alpha}\phi(x)$, with $\alpha$ constant. The energy of a field is given by the Hamiltonian
\begin{equation}
H_3 = \int d^3 x \left( {\dot \phi}^* {\dot \phi}  + \pa_i \phi^* \pa_i \phi  + V(\phi^*\phi) \right)\,.
\label{s2}
\end{equation}
The ground states are given by $\phi = v \ee^{\ii \th}/\sqrt{2}$ with $\th$ constant (and have zero energy). 

The spontaneous symmetry breaking of the global symmetry goes as follows. A field with minimal energy is chosen, say $\phi =v /\sqrt{2}$, and small perturbations around this field are considered. To describe these small perturbations, the following field parametrization is used
\begin{equation}
\phi = \exp\left(\fr{\ii\xi}{v}\right) \fr{v+\eta}{\sqrt{2}} = \fr{1}{\sqrt{2}} (v +\eta + \ii\xi + \dots)\,,
\label{s3}
\end{equation}
where $\eta$ and $\xi$ (as well as their derivatives) are considered to be small real fields. Substitution in ${\mathcal L}_3$ yields
\begin{equation}
{\mathcal L}_4 = \frac{1}{2} \left( \pa_\mu \eta \pa^\mu \eta - 2\mu^2 \eta^2 \right) + \frac{1}{2}  \pa_\mu \xi \pa^\mu \xi + \dots\,, 
\label{s4}
\end{equation}
where the dots represent higher order terms in the perturbation. The terms quadratic in the fields yield the field equations for the perturbations up to terms that are linear in the fields. As such, it is seen that the perturbations describe a real massive field $\eta$ with real mass $\sqrt{2}\mu$ and a real massless field $\xi$. The massless field is called the {\em Nambu-Goldstone field}.  

Hence, according to this account, the spontaneous breaking of a global symmetry amounts to the fact that there is a degenerate ground state, where the various ground states are connected by the global symmetry, and that perturbations are considered around a particular ground state. Because the ground states are not invariant under the global symmetry, the theory describing small perturbations around a particular ground state does not possess the original symmetry anymore and the global symmetry is broken. 

If the global symmetry is regarded as a gauge symmetry, then it is clear that this account is not manifestly gauge invariant due to the choice of ground state.  Since all the ground states are connected by the global symmetry, they actually all correspond to the same physical state. As we will discuss in the next section, a manifestly gauge invariant account is possible, according to which the global symmetry is never broken, and which leads to the same results.

\subsection{Alternative account without symmetry breaking}\label{ssbgsaa}
For this alternative account, the key is to consider the first equality in \eqref{s3} as just a field transformation, rather than as an ansatz for a perturbation expansion. By merely applying this field transformation, the symmetry is still there, albeit in a more complicated form. (In this respect, one often talks about {\em hidden symmetry} instead of symmetry breaking, see for example \cite{oraifeartaigh79}.) 

More explicitly, we can consider the field transformation $\phi =  \rho \ee^{\ii\theta}/\sqrt{2}$, where $\rho = \sqrt{2\phi^* \phi}$ and $\th = (1/2\ii) \ln(\phi/\phi^*)$. This transformation is only valid if $\phi \neq 0$ everywhere. Hence, the space of possible field configurations should be accordingly restricted. In terms of the new fields, the ground states are given by $\rho = v = \mu/\lambda$ and $\theta$ constant. We can now perform a further field transformation $\rho = v + \eta$, with $\eta > -v$, and $\theta = \xi/v$. The field $\eta$ (as well as $\rho$) is invariant under global $U(1)$ transformations, while the field $\xi$ transforms as $\xi \to \xi + \alpha$, with $\alpha$ constant. 

The Lagrangian density for the new variables is 
\begin{equation}
{\mathcal L}_5  = \frac{1}{2} \left(\pa_\mu \eta \pa^\mu \eta - 2\mu^2 \eta^2 \right) + \frac{1}{2}  \pa_\mu \xi \pa^\mu \xi - \lambda v \eta^3 - \frac{\lambda}{4} \eta^4+ \frac{1}{v}\eta \pa_\mu \xi \pa^\mu \xi + \frac{1}{2v^2}\eta^2 \pa_\mu \xi \pa^\mu \xi  \,.
\label{s7}
\end{equation}
The first two terms are the same as before and indicate that the field $\eta$ has mass ${\sqrt{2}}\mu$ and that the field $\xi$ is massless. The other terms represent interactions between the fields. Restricting the field $\xi$ to be small would violate the symmetry. However, instead, we can consider merely $\partial_\mu \xi$ to be small, which does not violate the symmetry. So if both $\eta$ and $\partial_\mu \xi$ are small, the effective Lagrangian is
\begin{equation}
{\mathcal L}_6 = \frac{1}{2} \left( \pa_\mu \eta \pa^\mu \eta - 2\mu^2 \eta^2 \right) + \frac{1}{2}  \pa_\mu \xi \pa^\mu \xi + \dots\,, 
\label{s8}
\end{equation}
which has the same form as before. The only difference is that $ \xi$ is not assumed to be small. 

So if we compare this account to the one above, the global symmetry is still present and the ground state is still degenerate. However, instead of restricting the set of solutions by considering perturbations around a particular ground state, they were restricted using only invariant conditions. Hence the symmetry was never violated or spontaneously broken. Yet, the resulting physics, namely the appearance of two real fields, one massive and one massless, is the same.

The theory can also be expressed explicitly in terms of gauge independent variables. In order to do this, note that the field equations corresponding to the Lagrangian density ${\mathcal L}_5 $ only depend on $\pa_\mu \xi$. Hence, one can replace $\pa_\mu \xi$ by a new field $B_\mu$ in the field equations and add the integrability condition $\pa_\mu B_\nu - \pa_\nu B_\mu = 0$ (on the level of the Lagrangian, this equation will need to be imposed by  means of Lagrangian multipliers). As such, up to linear terms, the field equations become: 
\begin{equation}
\square \eta + 2 \mu^2 \eta = 0 \,, \qquad \pa_\mu B^\mu = 0\,, \qquad  \pa_\mu B_\nu - \pa_\nu B_\mu = 0 
\label{s9}
\end{equation}
(the last two field equations are the same as those for a free electromagnetic field in the Lorentz gauge, with vanishing field tensor). If we had eliminated the global $U(1)$ symmetry like this from the start, then we would have found that there is a unique ground state and we would never have met the criteria for spontaneous symmetry breaking. Yet, for small fields $\eta$ and $B_\mu$, the same physics would have been obtained.

\section{Spontaneous symmetry breaking of a local $U(1)$ symmetry and the Higgs mechanism}\label{ssbls}
\subsection{Usual account}\label{ssblssa}
Consider the Lagrangian density given by 
\begin{equation}
{\mathcal L}_7  = (D_{\mu} \phi)^* D^{\mu} \phi  - V(\phi^*\phi)   - \frac{1}{4} F^{\mu \nu}F_{\mu \nu} \,,
\label{s10}
\end{equation}
where $V$ is the Higgs potential as given in \eqref{s1.1}. This theory has the local $U(1)$ symmetry given by \eqref{t6}. 

The corresponding energy is 
\begin{equation}
H_7 = \int d^3 x \left( (D_0 \phi)^* D_0 \phi + (D_i \phi)^* D_i \phi + V(\phi^*\phi) + \frac{1}{2}F_{0i}F_{0i} + \frac{1}{4}F_{ij}F_{ij}\right)\,.
\label{s11}
\end{equation}
In this case, the states with minimal energy are given by $\phi= v \ee^{\ii \th}/\sqrt{2}$ and $A_\mu=-\pa_\mu \th /e $, where $\th$ is an arbitrary space-time dependent function. 

Spontaneous breaking of the local $U(1)$ symmetry amounts to choosing a ground state, say $\phi= v /\sqrt{2},$ $A_\mu = 0$, and considering perturbations around it. The perturbations for the vector potential can be described by $A_\mu$ itself and perturbations for the scalar field are described by small real fields $\eta$ and $\xi$, defined by
\begin{equation}
\phi = \exp\left(\fr{\ii\xi}{v}\right) \fr{v+\eta}{\sqrt{2}} = \fr{1}{\sqrt{2}} (v +\eta + \ii\xi + \dots)\,.
\label{s12}
\end{equation}
Substituting the latter expansion into ${\mathcal L}_7$ leads to
\begin{equation}
{\mathcal L}_8  = \frac{1}{2} \left(\pa_\mu \eta \pa^\mu \eta - 2\mu^2 \eta^2 \right) + \frac{1}{2}  \pa_\mu \xi \pa^\mu \xi + evA_\mu\pa^\mu \xi + \frac{1}{2} e^2v^2 A_\mu A^\mu - \frac{1}{4} F^{\mu \nu}F_{\mu \nu} +\dots \,.
\label{s13}
\end{equation}
It is seen that the perturbations describe a real massive field $\eta$. The term $evA_\mu\pa^\mu \xi$ implies that the masses of the fields $\xi$ and $A_\mu$ are not immediately clear. However, by introducing the field 
\begin{equation}
B_\mu =  A_\mu   +  \frac{1}{ev}\partial_\mu  \xi\,, 
\label{s13.1}
\end{equation}
the Lagrangian density becomes
\begin{equation}
{\mathcal L}_9  = \frac{1}{2} \left(\pa_\mu \eta \pa^\mu \eta - 2\mu^2 \eta^2 \right) + \frac{1}{2} e^2v^2 B_\mu B^\mu - \frac{1}{4} B^{\mu \nu}B_{\mu \nu} +\dots \,,
\label{s15}
\end{equation}
where $B_{\mu\nu}=\pa_\mu B_\nu - \pa_\nu B_\mu$. The field $B_\mu$, which combines the vector potential with the Nambu-Goldstone boson, has mass $ev$. 

The field transformation \eqref{s13.1} could also be understood as part of a transformation \eqref{t6} with $\alpha = -\xi/v$:
\begin{equation}
\phi \to {\textrm e}^{-\ii \xi /v} \phi = \fr{1}{\sqrt{2}} (v +\eta) \,, \quad  A_{\mu} \to   A_\mu  - \frac{1}{e}\partial_\mu \left( - \fr{\xi}{v} \right)= A_\mu   +  \frac{1}{ev}\partial_\mu  \xi = B_\mu \,. 
\label{s14}
\end{equation}
This transformation projects the fields to the unitary gauge (given by $\phi=|\phi|$). This gauge could also be employed from the start, that is, before small perturbations are considered. In doing so, the ground state would be uniquely given by $v/\sqrt{2}$ and no Nambu-Goldstone boson would appear. For small perturbations around the ground state, the Lagrangian density \eqref{s15} would be obtained.  

In any case, these treatments are not manifestly gauge invariant. 

\subsection{Account in terms of gauge invariant variables}\label{ssblsaa}
Let us now turn to Higgs' manifestly gauge invariant treatment \cite{higgs66} (which can also be found in \cite{rubakov02}). A field transformation $(\phi,A_\mu) \to (\rho, \theta, B_\mu)$ is performed, defined by $\phi = \rho \ee^{\ii\theta}/ \sqrt{2}$, $B_\mu = A_\mu + \pa_\mu \theta/e$ $(=  - j_\mu/2e^2\phi^*\phi)$. The fields $\rho$ and $B_\mu$ are gauge invariant, while the field $\theta$ transforms as $\theta \to \theta + \alpha$ and is hence a pure gauge variable. Expressing ${\mathcal L}_4$ in terms of the new variables yields
\begin{equation}
{\mathcal L}_{10}  = \frac{1}{2} (\pa_\mu + \ii e B_\mu)^* \rho (\pa^\mu + \ii e B^\mu)\rho  - V(\rho) - \frac{1}{4} B^{\mu \nu}B_{\mu \nu}\,,
\label{s17}
\end{equation}
where $V(\rho) = \lambda (\rho^2 - v^2)/4$ and $B_{\mu\nu}=\pa_\mu B_\nu - \pa_\nu B_\mu$. It only depends on the gauge invariant variables $\rho$ and $B_\mu$. The gauge variable $\theta$ can be ignored from now on. 

Note that if instead we would employ the unitary gauge, then this would simply put $\theta=0$. As such, there is a close connection between this choice of gauge invariant variables and the unitary gauge. We spell this out in more detail in the following section.

The corresponding energy is given by
\begin{multline}
H_{10} = \int d^3 x \Big( (\pa_0 + \ii e B_0 )^* \rho (\pa_0 + \ii e B_0 ) \rho  + (\pa_i - \ii e B_i )^* \rho (\pa_i - \ii e B_i )\rho  + V(\rho) \\
+ \frac{1}{2}B_{0i}B_{0i} + \frac{1}{4}B_{ij}B_{ij}\Big)\,.
\label{s18}
\end{multline}
There is a unique ground state now, given by $B_\mu =0$, $\rho = v$. By performing the transformation $\rho = v + \eta$, the ground state becomes $\eta = 0$. Perturbations around it are described by the Lagrangian density 
\begin{equation}
{\mathcal L}_{11}  = \frac{1}{2} \left(\pa_\mu \eta \pa^\mu \eta - 2\mu^2 \eta^2 \right) + \frac{1}{2} e^2v^2 B_\mu B^\mu - \frac{1}{4} B^{\mu \nu}B_{\mu \nu} +\dots \,.
\label{s20}
\end{equation}
This is exactly the Lagrangian density \eqref{s15} that was obtained earlier. But this time only gauge independent variables were employed.

Note that the above analysis is only valid for fields $\phi$ that are everywhere non-zero. In section \ref{ssbhamiltonian}, in the context of the Hamiltonian picture, the local $U(1)$ symmetry will be eliminated, up to a residual global $U(1)$ symmetry, by using a field transformation that is defined also for fields that are not necessarily non-zero everywhere.

\section{Gauge fixing and gauge invariant variables}\label{gaugefixing}
In the previous section, it was noted that the formulation in terms of gauge invariant variables is exactly the same as the one that would be obtained by imposing the unitary gauge. For other (suitable) gauges, one can similarly find a field transformation that separates gauge independent variables from gauge degrees of freedom, such that the gauge invariant variables resemble the gauge fixed variables. We want to sketch this in the current section (see also \cite{ilderton10,salisbury09}).{\footnote{We will mention analogous results in the Hamiltonian picture, in section \ref{underdeterminationhamiltonian}, where this can be spelled out a bit more precisely.}} 

Consider a gauge fixing of the form $F(A_\mu,\phi)=0$. Usually, the gauge fixing is required to uniquely fix the gauge and to be attainable, meaning that there exists a function $\alpha^F(A_\mu,\phi)$ such that for all fields $A_\mu,\phi$, the fields 
\begin{equation}
\phi^F= {\textrm e}^{\ii \alpha^F} \phi \,, \qquad     A^F_\mu = A_\mu -  \frac{1}{e}\partial_\mu  \alpha^F \,, 
\label{f1}
\end{equation}
satisfy the gauge, that is, $F(A^F_\mu,\phi^F) \equiv 0$. Equation \eqref{f1} can then be regarded as defining a field transformation from the field variables $(\phi,A_\mu)$ to $(\phi^F,A^F_\mu,\alpha^F)$, with inverse transformation
\begin{equation}
\phi = {\textrm e}^{- \ii \alpha^F} \phi^F \,, \qquad     A_\mu = A^F_\mu +  \frac{1}{e}\partial_\mu  \alpha^F \,. 
\label{f2}
\end{equation}
Since fields that are connected by a gauge transformation are projected to the same fields $\phi^F,A^F_\mu$, those variables are gauge invariant. The variable $\alpha^F$ transforms as $\alpha^F \to \alpha^F - \alpha$ and is pure gauge. So the field transformation yields a separation of gauge invariant variables and pure gauge variables. In terms of the new variables, a gauge fixing just amounts to fixing the variable $\alpha^F$. As such, if one wants to take seriously the ontology suggested by gauge fixing, then there is very little difference compared to an ontology in terms of gauge independent variables. In the latter, the ontology is given by $\phi^F,A^F_\mu$. In the former, it is given by same fields together with an additional field $\alpha^F$, which is then a fixed function. 

Let us now apply this to the unitary gauge. This gauge is given by $\phi = |\phi|$ or, writing $\phi = \rho \ee^{\ii\theta}/ \sqrt{2}$, by $\th=0$. Assuming $\phi \neq 0$ over all space-time, this uniquely fixes the gauge. In this case $\al^U(A_\mu,\rho,\th)= - \th$, so that 
\begin{equation}
\phi^U = \frac{1}{\sqrt{2}}  \rho \,, \qquad A^U_\mu = A_\mu + \frac{1}{e} \pa_\mu \th \,.
\label{f3}
\end{equation}
The corresponding field transformation to the variables $\phi^U,A^U_\mu,\al^U$ corresponds to the one that what was applied in the previous section to separate gauge degrees of freedom from gauge independent ones. 

Some common gauge conditions do not completely fix the gauge. One example is the Coulomb gauge, which reads $\pa_i A_i = 0$. This condition still allows for local $U(1)$ transformations \eqref{t6} with $\nabla^2 \alpha = 0$. In sections \ref{example} and \ref{ssbreconsidered}, we will consider a natural choice of boundary conditions for the fields, for which the only gauge transformations that preserve those boundary conditions and which satisfy $\nabla^2 \alpha = 0$, are those for which $\alpha$ is constant. The fields can be projected to those satisfying the Coulomb gauge using the function $\alpha^C = -e \nabla^{-2 } \pa_i A_i $, where $\nabla^{-2 } f({\bf x})= - \int d^3 y f({\bf y})/4\pi|{\bf x}-{\bf y}|$ (which is well-defined for the given boundary conditions), so that
\begin{equation}
\phi^C = \ee^{- \ii e \nabla^{-2} \partial_iA_i}  \phi \,, \quad A^C_i = A_i -  \partial_i \frac{1}{\nabla^2} \partial_jA_j \,, \quad A^C_0 = A_0 + \pa_0   \frac{1}{\nabla^2} \partial_j A_j \,.
\label{f4}
\end{equation}
The fields $A^C_0$ and $A^C_i$ (which is the transverse part of the vector potential) are invariant under local $U(1)$ transformations (that preserve the assumed boundary conditions), while the field $\phi^C$ may still pick up a global phase. This implies that after ignoring the field $\alpha^C$ the theory will still have a residual global $U(1)$ symmetry. 

This choice of variables is very natural when the alternative notion of gauge is employed that is to be considered in the following sections. Namely, according to that notion, the residual global $U(1)$ symmetry will not constitute a gauge symmetry. These variables are also most conveniently handled in the Hamiltonian picture. We will therefore reconsider them in section \ref{ssbhamiltonian}, where we discuss the Hamiltonian picture and reduced phase space, and postpone a discussion of the Higgs mechanism until then.

\section{Indeterminism and gauge}\label{gaugesymindet}
\subsection{A notion of gauge symmetry}
As was already noted before, the presence of a local symmetry implies the breakdown of determinism. The equations of motion will admit different solutions with the same initial data. The initial data are the fields and their time derivatives at a particular time. This indeterminism may serve as a basis to define gauge symmetry (see \cite{earman04b,belot08a} for good discussions). According to this definition, a gauge transformation maps solutions of the equations of motion to solutions and preserves the initial data at a particular time, or is a combination of such maps. The time at which the initial data are considered may differ for different gauge transformations. A combination of maps that preserve initial data at some time does not necessarily preserve initial data itself, but it should also be considered a gauge transformation, because gauge transformations should define an equivalence relation (transitivity requires that if solutions 1 and 2 are gauge equivalent, and 2 and 3 are, then solutions 1 and 3 should be). The gauge invariant degrees of freedom are then those variables whose time evolution is deterministic.

Note that (non-trivial) global symmetries do not preserve some initial data. Therefore if a global symmetry group is not a subgroup of a local symmetry group, it will not contain (non-trivial) gauge transformations. On the other hand, in the case of a local symmetry group, Noether's second theorem implies the existence of a non-trivial gauge group. As will be illustrated with the examples below, this group depends very much on the boundary conditions that are imposed on the fields and does not necessarily correspond to the local symmetry group itself. 

This definition of gauge is customary within the context of constrained dynamics \cite{dirac64,hanson76,sundermeyer82,gitman90,henneaux91}, which concerns the Hamiltonian formulation of theories with a singular Lagrangian (such as those possessing a local symmetry).{\footnote{Note that in these texts, gauge transformations are often viewed as point transformations at a certain time, rather than as transformations of solutions. See \cite{pons05} for the relation between these notions.}} As we will discuss in detail in section \ref{hamiltonianpicture}, the indeterminism and hence the gauge freedom becomes explicit in the Hamiltonian formulation. For now, we restrict our attention to the Lagrangian picture.

\subsection{Example: scalar electrodynamics}\label{example}
Consider again the Lagrangian density ${\mathcal L}_2$, which describes scalar electrodynamics. Denote by ${\mathcal G}^b$ the group of local $U(1)$ transformations that preserve some boundary conditions of the fields $\phi$ and $A_\mu$. Further, denote by ${\mathcal G}^g$ the group of gauge transformations, that is, the group generated by elements $g \in {\mathcal G}^b$ that preserve some initial data, which means that there exists a time $t_0$ such that
\begin{equation}
g|_{t=t_0} = 1, \quad \pa_\mu g|_{t=t_0}  = \pa^2_0 g|_{t=t_0} =0 \,. 
\label{t8}
\end{equation}
The group ${\mathcal G}^g$ forms a normal subgroup of ${\mathcal G}^b$. By eliminating the gauge freedom, the set of solutions ${\mathcal S}$ is reduced to ${\mathcal S}/{\mathcal G}^g$. There might be a residual symmetry group acting on this space, namely ${\mathcal G}^b/{\mathcal G}^g$. This is then considered to be the group of physical symmetries.

If no boundary conditions are assumed, then ${\mathcal G}^b$ contains all the local $U(1)$ transformations and the gauge group ${\mathcal G}^g$ equals ${\mathcal G}^b$. To see this, consider $g= \ee^{\ii \alpha} \in {\mathcal G}^b$. Then $g=g_1g_2$, where $g_1=\ee^{\ii \alpha f}$ and $g_2=\ee^{\ii \alpha(1-f)}$, with $f$ a smooth function with $f=0$ for $t \leqslant 0$ and $f=1$ for $t \geqslant 1$. Because $g_1$ and $g_2$ are given by the identity for respectively $t \leqslant 0$ and $t \geqslant 1$, they are both elements of $ {\mathcal G}^g$. As such $g \in {\mathcal G}^g$.

However, often certain boundary conditions are imposed. This is done, for example, to ensure finiteness of the energy and action, and to ensure that the variation of the action is well defined (that is, that partial integrations can be performed in the variation without the appearance of boundary terms), see for example \cite{lusanna95,lusanna96a}. A natural choice of boundary conditions, which is sufficient to meet these requirements, is given by{\footnote{As is customary, $F = o(r^{-n})$ means that there exists a $c>0$ and an $r_0$, such that $F(r,\vartheta,\varphi) < c/r^n$ for $r> r_0$, with $(\vartheta,\varphi)$ the spherical angles.}} 
\begin{equation}
A_0 = o(r^{-1})\,,\quad A_i= o(r^{-2})\,,\quad \partial_\mu A_\nu = o(r^{-2}) \,,\quad \phi = o(r^{-2})\,,\quad \partial_\mu \phi = o(r^{-2}) \,,
\label{t9}
\end{equation}
as $r=|{\bf x}| \to \infty$. 
 
The local $U(1)$ transformations that preserve those boundary conditions are of the form $g=\ee^{\ii \alpha}$, where $\alpha$ goes to a constant (that is, independent of the angular coordinates and of time) sufficiently fast, as $r \to \infty$. In other words, $g$ should go to an element of $U(1)$ at spatial infinity, at an appropriate rate. The group of such transformations, which was previously called ${\mathcal G}^b$, is also called ${\mathcal G}^c$.{\footnote{We adopt this notation from \cite{balachandran94} where similar groups and similar considerations figure.}} The gauge group ${\mathcal G}^g$ is now given by ${\mathcal G}^\infty$, which is the subgroup of ${\mathcal G}^c$ of transformations that go to the identity at spatial infinity. This is shown similarly as for the case in which there are no boundary conditions. In this case, there is a non-trivial residual group of physical symmetries acting on the space ${\mathcal S}/{\mathcal G}^g$, namely ${\mathcal G}^b/{\mathcal G}^g = {\mathcal G}^c/{\mathcal G}^\infty \cong U(1)$.

Note that different boundary conditions could have been imposed to ensure the above mentioned requirements (in particular, weaker ones). These could potentially lead to different results concerning the gauge group and the related physical symmetry group. We present a simple example in the next section, for the case of the Higgs potential. This seems to be an unattractive feature of this notion of gauge symmetry.

\subsection{Spontaneous symmetry breaking reconsidered}\label{ssbreconsidered}
If gauge symmetry is linked to a failure of determinism, then there is no gauge symmetry in the case of a global symmetry. Solutions connected by a global $U(1)$ transformation, which are arguably observationally indistinguishable, are then regarded as physically distinct. As such, one can adopt the usual account of spontaneous symmetry breaking.

In the case of a local symmetry, the boundary conditions on the fields should be considered in order to establish the gauge group. A suitable set of boundary conditions is given by 
\begin{equation}
A_\mu= o(r^{-2})\,,\quad \partial_\mu A_\nu = o(r^{-2}) \,,\quad \phi =\frac{1}{\sqrt{2}}v \ee^{\ii{\bar \theta}} + o(r^{-2}) \,,\quad \partial_\mu \phi = o(r^{-2}) \,,
\label{t10}
\end{equation}
as $r\to \infty$, where ${\bar \theta}$ is an arbitrary real constant. So the scalar field should go to a minimum of the potential at infinity, at an appropriate rate. Similarly as in the case of ordinary scalar electrodynamics, these boundary conditions are preserved by the local $U(1)$ transformations $g=\ee^{\ii \alpha}$, for which $\alpha$ goes to a constant sufficiently fast at spatial infinity. We again denote this group by ${\mathcal G}^c$. The gauge group is given by ${\mathcal G}^g={\mathcal G}^\infty$ of transformations that go to the identity at spatial infinity. Elimination of the gauge degrees of freedom leaves a residual $U(1)$ symmetry. 

The gauge symmetry can be eliminated in a similar way as in section \ref{ssbgsaa}. In that section, all the local symmetries were considered gauge symmetries and as such the phase $\theta$ of the scalar field was identified as the gauge degree of freedom. With the current notion of gauge, the field $\theta$ is not completely gauge. Namely, its value at spatial infinity $\theta_\infty$ must now be considered as a physical degree of freedom. It remains constant over time, by the mere choice of boundary conditions, and hence its time evolution is trivially deterministic. This degree of freedom carries the residual $U(1)$ symmetry.

Another way to eliminate the gauge freedom is to use the transformation defined in \eqref{f4}, so that the gauge invariant degrees of freedom are the transverse part of the vector potential and the dressed scalar field. We will do this explicitly in the Hamiltonian picture, by passing to reduced phase space, in section \ref{ssbhamiltonian}. 

Alternatively, one could impose the boundary conditions \eqref{t10}, but with ${\bar \theta}$ fixed, say ${\bar \theta} = 0$, so that
\begin{equation}
 \phi =\frac{1}{\sqrt{2}}v  + o(r^{-2}) \,,\quad {\textrm{ as }}  r \to \infty \,.
\label{t11}
\end{equation}
In this case, the group of local $U(1)$ transformations that preserves the boundary conditions is given by ${\mathcal G}^\infty$. Since this group coincides with the group of gauge transformations, there is no residual group of physical symmetries.

\subsection{Excursion to Yang-Mills theories}\label{etymt}
While we will not discuss spontaneous symmetry breaking for Yang-Mills theories, we want to briefly discuss some properties of the gauge group.{\footnote{Similar discussions can be found in \cite{balachandran91,balachandran94,lusanna95,giulini95b}.}} 

In Yang-Mills theories, the local symmetry group is given by smooth maps from space-time to a compact and connected Lie group $G$. The vector potential $A_\mu$, which takes values in the Lie algebra of $G$, transforms as 
\begin{equation}
A_{\mu}(x) \to  g(x) A_\mu(x)g(x)^{-1} -  \frac{\ii}{e} g(x)\partial_\mu  g(x)^{-1}
\label{t12}
\end{equation}
under the local symmetry. The matter fields transform according to some unitary representation of $G$. As before, appropriate fall off conditions could be assumed. For simplicity, we assume that they are such that they impose the restriction on the possible local transformations that $\lim_{r \to \infty} g = g_\infty(\varphi,\vartheta)$, where $g_\infty$ is a time independent function which may depend on the angular coordinates, and where the limit value is reached at some particular rate. 

Let ${\mathcal G}^b$ be the group of local symmetries that preserve those boundary conditions and ${\mathcal G}^\infty$ its normal subgroup of transformations that go to the identity at spatial infinity. Similarly as in the case of a local $U(1)$ symmetry, the symmetries $g \in {\mathcal G}^b$ that preserve the fields and their time derivatives at a time $t_0$ satisfy $g|_{t=t_0} = 1$, $\pa_\mu g|_{t=t_0}  = \pa^2_0 g|_{t=t_0} =0$. The gauge group ${\mathcal G}^g$ is generated by those elements. In this case, it is given by ${\mathcal G}^\infty_0$, the component of ${\mathcal G}^\infty$ continuously connected to the identity. This is shown in the appendix. As such, large transformations (that is, symmetry transformations that are not continuously connected to the identity transformation) are not considered to be gauge transformations. Eliminating the gauge group will result in a residual symmetry group isomorphic to ${\mathcal G}^b/{\mathcal G}^g = {\mathcal G}^b/{\mathcal G}^\infty_0$. 

We have that ${\mathcal G}^\infty/{\mathcal G}^\infty_0 \cong \pi_0({\mathcal G}^\infty) \cong \pi_0({\bar {\mathcal G}}^\infty)$, where ${\bar {\mathcal G}}^\infty$ is the group of smooth maps from ${\mathbb R}^3$ to $G$ that go to the identity at spatial infinity (see appendix). As in \cite{balachandran91,balachandran94}, we can then write $\pi_0({\bar {\mathcal G}}^\infty) \cong \pi_0({\tilde {\mathcal G}}^\infty)$, with ${\tilde {\mathcal G}}^\infty$ the group of smooth maps from $S^3$ to $G$ that map a fixed point on $S^3$ to the identity, where $S^3$ arises as the one point compactification of ${\mathbb R}^3$ by adding a point at spatial infinity. As such, $\pi_0({\bar {\mathcal G}}^\infty) \cong \pi_3(G)$. If $G$ is Abelian (for example $G=U(1)$), then $\pi_3(G) \cong \{1\}$ and hence ${\mathcal G}^\infty = {\mathcal G}^\infty_0$. In the case of a simple group (for example $G=SU(N)$, $N>1$), $\pi_3(G) \cong {\mathbb Z}$, so that ${\mathcal G}^\infty \neq {\mathcal G}^\infty_0$.

\subsection{Subsystems and gauge}\label{sorkin}
There is also a different possible interpretation of the boundary conditions and the groups ${\mathcal G}^b$ and ${\mathcal G}^\infty$, explained to us by Sorkin \cite{sorkinprivate}, which is worth presenting here. 

One could take the view that the fields effectively describe a bounded subsystem of the universe, rather than the universe itself \cite{friedman80a,faddeev82,giulini95a,giulini96}. As such, spatial infinity should not be regarded as representing the boundary {\em of} space, but rather as a boundary {\em in} space (and similarly for temporal infinity, although we will not consider the latter). As such, the boundary conditions can be regarded as an idealization of the fact that, the fields of the subsystem must still match up smoothly with those representing the environment, which are held fixed.

What is then the interpretation of groups like ${\mathcal G}^b$ and ${\mathcal G}^\infty$? ${\mathcal G}^b$ corresponds to what Sorkin prefers to call partial symmetry transformations (to distinguish them from local symmetry transformations).{\footnote{While such transformations often appear in the work of Sorkin and collaborators, see for example \cite{friedman80a,friedman83a,friedman83b}, the actual terminology does not seem to have made it into any of their papers.}} They are transformations of the subsystem which take the form of local transformations in the region occupied by the subsystem. Partial transformations may not correspond to actual local transformations when viewed as transformations over all space-time. A partial transformation can only be regarded as a local transformation, when it goes to the identity at the boundary and this at an appropriate rate such that the smoothness of the transformation is guaranteed (since local transformations must be smooth). So, in the idealization that takes the boundary to infinity, this group of transformations corresponds to ${\mathcal G}^\infty$. 

Taking the view that local transformations form gauge transformations, the group ${\mathcal G}^\infty$ contains the partial transformations that correspond to gauge transformations (and not ${\mathcal G}^\infty_0$ as in the previous section). The other partial transformations do not correspond to a local transformation and can hence be viewed as changing the physical state of the subsystem. However, the observational difference only arises when comparing the fields of the subsystem with those of the environment and not from within the subsystem itself. This is analogous to the familiar view one can take on, for example, translation and rotation symmetry in classical mechanics. While translations or rotations of the whole universe are unobservable, and hence may be regarded as gauge transformations, translations or rotations of (approximately isolated) subsystems relative to their environment will yield (in principle) observable differences.{\footnote{It is also in terms of such partial transformations that an analogon of Galileo's ship experiment could be conceived, not in terms of local transformations \cite{brading04,healey07,healey09}.}}

Let us illustrate this for scalar electrodynamics. Suppose a certain spatial region and fields which vanish at the boundary of the region. Transformations of the fields in the spatial region of the form $\phi \to \ee^{\ii \alpha} \phi$, $A_\mu \to A_\mu$, with $\alpha$ constant, form examples of partial transformations. If $\alpha$ is non-zero, then these partial transformations can not be regarded as a local $U(1)$ transformations.{\footnote{To find an observable difference corresponding to such a phase difference between system and environment, one could let part of the field of the subsystem interfere with part of the field of the environment. Provided both parts are non-zero, there will be a fringe shift in the charge density (which is an observable due to its invariance under local $U(1)$ transformations). In principle, the disturbance to the subsystem could be made arbitrarily small, by only letting a sufficiently small part of it interfere with the environment.}} The partial transformations that correspond to gauge transformations must go to the identity at the boundary. Elimination of the gauge freedom leads to a residual group of physical symmetries. As in section \ref{example}, this group will be isomorphic to $U(1)$. 

In the case the scalar field is non-zero at the boundary, partial transformations must go to the identity at the boundary in order to preserve the smoothness of the scalar field. As such, all partial transformations correspond to gauge transformations and there is no residual symmetry group. For example, the boundary conditions \eqref{t11}, which fix the scalar field to be a particular minimum of the Higgs potential at spatial infinity, could be understood as an idealization of such a situation (identifying the spatial boundary with spatial infinity). The boundary conditions \eqref{t10} do not fix the field at spatial infinity and hence cannot be understood as corresponding to boundary conditions for an actual subsystem. 

Balachandran {\em et al.}\ also consider the group ${\mathcal G}^\infty$ as the group of gauge transformations \cite{balachandran91}. However, their motivation seems to be different than the one expressed above. Their motivation to take ${\mathcal G}^\infty$ and not ${\mathcal G}^\infty_0$ as the group of gauge transformations is that one would need non-local observables to distinguish the fields that are connected by elements of ${\mathcal G}^\infty$ that are not in the same component. According to Balachandran {\em et al.}, it is then difficult to imagine a reasonable experiment that could determine the value of those observables \cite[p.\ 284]{balachandran91}. However, if these considerations were applied to subsystems, instead of to the whole universe, then this worry would probably disappear.

\section{Hamiltonian picture}\label{hamiltonianpicture}
\subsection{Indeterminism, gauge and reduced phase space}\label{underdeterminationhamiltonian}
When passing from the Lagrangian to the Hamiltonian picture, the existence of a local symmetry group leads to the appearance of arbitrary functions of time in the Hamiltonian. As such, different choices for those functions will lead to different time evolutions, even if the initial data are held fixed. As such, the breakdown of determinism becomes explicit.

The details of the Hamiltonian formulation can be found in \cite{dirac64,hanson76,sundermeyer82,gitman90,henneaux91}. For a conceptual overview see \cite{earman03a}. Here, we just recall the basic ingredients for a system described by a finite number of degrees of freedom. Given a Lagrangian $L(q,{\dot q})$, where $q=(q_1,\dots,q_N)$, the passage to the Hamiltonian formulation involves certain constraints $\chi_m(q,p)=0$, $m=1,\dots,M$ on phase space, which can be classified as either {\em first} or {\em second class}. A constraint $\chi_m$ is first class if $\{\chi_m,\chi_{m'}\} \approx 0$, for $m'=1,\dots,M$, where $\{.,.\}$ denotes the Poisson bracket and $\approx$ denotes the {\em weak equality}, that is, the equality holds after imposing the constraints. Otherwise it is a second class constraint. 

In the case there are only first class constraints, the dynamics is determined by the extended Hamiltonian{\footnote{There actually exist examples where not all the first class constraints should be included in the Hamiltonian. Those constraints then do not lead to an underdetermination of the equations of motion. These examples form counter examples to the ``Dirac conjecture'' and could be avoided by further regularity conditions (see \cite{henneaux91} for a detailed discussion). However, these issues do not concern us here.}}
\begin{equation}
H_E = H_C + \sum^M_{m=1} u_m \chi_m \,,
\label{u1}
\end{equation}
where $H_C = \sum^N_{n=1} p_n {\dot q}_n - L$ is the canonical Hamiltonian (expressed in terms of the canonical variables $(q,p)$) and the $u_m$ are arbitrary functions of the canonical variables and of time. The equation of motion for a dynamical variable $F(q,p)$ is given by 
\begin{equation}
{\dot F} \approx \{F,H_E\} \approx \{F,H_C\}  + \sum^M_{m=1} u_m \{F , \chi_m\} \,.
\label{u2}
\end{equation}
Hence, if $\{F , \chi_m\} \not\approx 0$, then the time evolution of $F$ will depend on the arbitrary function $u_m$. This is where gauge comes into play at the Hamiltonian level. Gauge independent variables, which are those variables whose time evolution is uniquely determined by the equations of motion and the initial data, should have weakly vanishing Poisson brackets with the constraints.

In the case of second class constraints, the Hamiltonian is again different from the canonical Hamiltonian, but it does not depend on arbitrary functions of time. Therefore there is no gauge freedom. 

In either case, the theory can be reformulated in terms of unconstrained canonical pairs, corresponding to what is called {\em reduced phase space}. For a theory with first class constraints, those unconstrained pairs form a complete set of gauge independent variables. The reduced phase space is obtained by performing a canonical transformation $(q,p) \to (q',p',q'',p'')$, where the $p'$ and $p''$ are the momenta conjugate to respectively $q'$ and $q''$, such that, in terms of the new variables, the constraints are $p''=0$. Such a transformation can be performed at least locally, see for example \cite[pp.\ 36-45]{gitman90} or \cite{maskawa76}. The equation of motion for a function $F(q',p')$ is of the usual form:
\begin{equation}
\dot{F} = \{F,H_P\}\,,
\label{u3}
\end{equation}
where
\begin{equation}
H_P(q',p') = H(q',p',q'',p'')\big|_{p''=0}
\label{u4}
\end{equation}
and where $H(q',p',q'',p'')$ is the Hamiltonian that is obtained from $H(q,p)$ by the canonical transformation. The Hamiltonian $H_P$, which does not depend on the variables $q''$, is called the {\em physical Hamiltonian}. The motion of the variables $q''$ is completely undetermined, which makes them gauge variables. As such, the variables $q''$, together with the variables $p''$ (which are constrained to be zero), can be ignored in the description of the system. It can further be shown that the physical degrees of freedom $q'$ and $p'$ are unique up to a canonical transformation \cite[p.\ 40]{gitman90}. 

In the case of second class constraints, one can similarly pass to reduced phase space by performing a canonical transformation $(q,p) \to (q',p',q'',p'')$, such that, in terms of the new variables, the constraints read $q''=p''=0$, see \cite[p.\ 27-35]{gitman90} and \cite{maskawa76}. The equation of motion for a function $F(q',p')$ of the unconstrained variables is
\begin{equation}
\dot{F} = \{F,H_P\} \,,
\label{u5}
\end{equation}
where 
\begin{equation}
H_P(q',p') = H(q',p',q'',p'')\big|_{q''=p''=0}
\label{u6}
\end{equation}
and where $H(q',p',q'',p'')$ is the Hamiltonian that is obtained from $H(q,p)$ by the canonical transformation. The unconstrained degrees of freedom $q'$ and $p'$ are unique up to a canonical transformation \cite[p.\ 32]{gitman90}.

So for a gauge theory, determinism is restored by passing to reduced phase space. Alternatively, one could employ a gauge fixing, that is, impose by hand further constraints on the phase space variables. A suitable gauge fixing should turn the complete set of constraints second class. So while the reduced phase space approach ignores the variables $q''$, whose time evolution is arbitrary, a gauge fixing amounts to fixing $q''$ in terms of the other coordinates. Either way, the dynamics of the variables $(q',p')$ remains the same. As such, as was already noted in section \ref{gaugefixing}, there is hardly any difference between an ontology in terms of gauge fixed variables and one in terms of gauge invariant variables.

Note that if one wants to eliminate the gauge freedom at the Lagrangian level, one could first pass to the Hamiltonian picture and eliminate the gauge freedom there, by passing to reduced phase, and then perform a Legendre transformation to move back to the Lagrangian picture.

Much of this formalism carries over to field theories, though with certain complications, see for example \cite{sundermeyer82}. It is unclear to us whether similar uniqueness results exist concerning reduced phase space.

\subsection{The Higgs mechanism in reduced phase space}\label{ssbhamiltonian}
Consider again the Lagrangian ${\mathcal L}_7$ for the Abelian Higgs model. In section \ref{ssbreconsidered}, we found that for a natural choice of boundary conditions, given by  \eqref{t10}, the gauge group is given by the group ${\mathcal G}^\infty$ of local $U(1)$ transformations that go to the identity at spatial infinity, at an appropriate rate. We also found that the elimination of the gauge freedom must yield a residual $U(1)$ symmetry. We illustrate this here by passing to reduced phase space.

One can pass to reduced phase space by implementing Higgs' field transformation of section \ref{ssblsaa}, which separates the gauge independent degrees of freedom from the gauge degrees of freedom, as a canonical transformation.  This was done by Lusanna and Valtancoli in \cite{lusanna96a}. As discussed in section \ref{ssbreconsidered}, the variable $\theta_\infty$ should also be kept as a physical degree of freedom, as it carries the residual $U(1)$ symmetry. But since $\theta_\infty$ is constant the spontaneous symmetry breaking of this residual symmetry is trivial.

The disadvantage of this approach is that it employs a polar decomposition of the scalar field, which is only valid when the scalar field is nowhere zero. Therefore, we present a different way to pass to reduced phase space, which does not pose such a restriction. The variables that will parametrize the reduced phase space are the transverse part of the vector potential and the dressed scalar field, which were already encountered in section \ref{gaugefixing}, in connection with the Coulomb gauge. They were originally employed by Dirac \cite{dirac55} to parametrize the reduced phase space in the context of scalar electrodynamics. Lusanna and Valtancoli actually also considered these variables, but did not consider the Higgs mechanism in terms of them. We will do this in the present section (see also \cite{struyve10} where we presented this account in a different context).

\subsubsection{Hamiltonian picture and reduced phase space}\label{hphm}
The details of the Hamiltonian picture and reduced phase space corresponding to the Lagrangian density ${\mathcal L}_7$ can be found in \cite[pp.\ 113-127]{gitman90} and \cite{lusanna96a,struyve10}. Here is an outline. 

Denoting the field conjugate to $\phi$ by $\Pi_{\phi}$, that is, $\Pi_{\phi}(x)= \pa {\mathcal L}_7/\pa {\dot \phi}(x)$, and similarly for the other fields, the conjugate momenta are given by{\footnote{The fields $\phi$ and $\phi^*$ are treated as independent. Equivalently, real fields $\phi_1$ and $\phi_2$ could be introduced, with $\phi = \phi_1 + \ii \phi_2$.}}
\begin{equation}
\Pi_{\phi} = D^*_0 \phi^{*}\,, \quad \Pi_{\phi^*} = D_0 \phi \,, \quad \Pi_{A_0} =  0\,, \quad \Pi_{A_i}  = ({\dot A}_i +\partial_i A_0 ) \,.
\label{h1}
\end{equation}
The canonical Hamiltonian is 
\begin{equation}
H_C =   \int d^3 x \left(\Pi_{\phi^*} \Pi_{\phi} + \left(D_{i} \phi \right)^{*}  D_{i} \phi  + V(\phi^*\phi)+  \frac{1}{2} \Pi_{A_i}\Pi_{A_i} + \frac{1}{4} F_{ij}F_{ij} + A_0\left( \partial_i \Pi_{A_i} + j_0 \right) \right) \,.
\label{h2}
\end{equation}
There are two first class constraints $\chi_1 = \chi_2 = 0$, given by 
\begin{equation}
\chi_{1} = \Pi_{A_0}\,, \quad \chi_{2} = \partial_i \Pi_{A_i} + j_0 =  \partial_i \Pi_{A_i} + \ii e\left(\phi^*  \Pi_{\phi^*} - \phi \Pi_{\phi} \right) \,.
\label{h3}
\end{equation}
The latter constraint is recognized as Gauss' law.

As discussed in section \ref{underdeterminationhamiltonian}, in order to pass to reduced phase space, a canonical transformation should be performed such that in terms of the new variables the constraints are given by some of the conjugate momenta. In this case, the constraint $\chi_1 = \Pi_{A_0}$ is already given by a field momentum, so that $A_0$ is readily identified as a gauge degree of freedom. In order to bring the Gauss constraint $\chi_2=0$ into the form of a field momentum, the following transformation can be performed:
\begin{align}
{\bar \phi} &= \ee^{- \ii e \frac{\partial_i}{\nabla^2} A_i}  \phi \,, & \Pi_{{\bar \phi}} &= \ee^{\ii e \frac{\partial_i}{\nabla^2} A_i} \Pi_{\phi} \,, \nonumber\\
{\bar \phi}^*  &= \ee^{\ii e \frac{\partial_i}{\nabla^2} A_i} \phi^* \,, & \Pi_{{\bar \phi}^*} &= \ee^{-\ii e \frac{\partial_i}{\nabla^2} A_i} \Pi_{\phi^*} \,, \nonumber\\
A^T_i &= \left(\delta_{ij} - \frac{\partial_i \partial_j}{\nabla^2} \right)A_j \,, & \Pi_{A^T_i} &= \left(\delta_{ij} - \frac{\partial_i \partial_j}{\nabla^2} \right)\Pi_{A_j} \,, \nonumber\\
A^L_i &= \frac{\partial_i \partial_j}{\nabla^2} A_j \,, & \Pi_{A^L_i} &= \frac{\partial_i }{\nabla^2} ( \partial_j\Pi_{A_j}  + j_0)\,.
\label{h4}
\end{align}
(Note that this transformation is well defined because of the choice of boundary conditions.) In terms of the new variables, the Gauss constraint is given by $\Pi_{A^L_i}=0$. So a complete set of gauge independent variables is given by ${\bar \phi},\Pi_{{\bar \phi}},{\bar \phi}^*,\Pi_{{\bar \phi}^*},A^T_i,\Pi_{A^T_i}$ \cite{dirac55}.

Note that the variables $\{{\bar \phi},\Pi_{{\bar \phi}}\}$ and $\{{\bar \phi}^*,\Pi_{{\bar \phi}^*}\}$ form canonical pairs. But the pairs $\{A^T_i,\Pi_{A^T_i}\}$ and $\{A^L_i,\Pi_{A^L_i}\}$ do not. Their Poisson brackets are respectively given by the transverse delta function $\left(\delta_{ij} - \partial_i \partial_j/\nabla^2 \right)\delta({\bf x} - {\bf y})$ and the longitudinal delta function $\left( \partial_i \partial_j/\nabla^2\right) \delta({\bf x} - {\bf y})$. So strictly speaking this transformation does not form a canonical transformation. However, by passing to Fourier space, these variables can easily be expressed in terms of canonical pairs (see for example \cite{struyve10}).

The Hamiltonian for the gauge independent variables is given by
\begin{equation}
H_P = \int d^3 x \bigg(\Pi_{{\bar \phi}^*} \Pi_{{\bar \phi}} + \left(D^T_{i} {\bar \phi}  \right)^{*}  D^T_{i} {\bar \phi}  +  V({\bar \phi}^*{\bar \phi}) -\frac{1}{2} {\bar j}_0  \frac{1}{\nabla^2} {\bar j}_0  +  \frac{1}{2} \Pi_{A^T_i}\Pi_{A^T_i} - \frac{1}{2} A^T_i \nabla^2 A^T_i \bigg)\,, 
\label{h5}
\end{equation}
where $D^T_{i} = \partial_{i} - \ii e A^T_i$ and ${\bar j}_0 = \ii e\left({\bar \phi}^*   \Pi_{{\bar \phi}^*} - {\bar \phi}  \Pi_{{\bar \phi} } \right)$ is the charge density expressed in terms of these variables. The term containing ${\bar j}_0$ is the Coulomb energy. Note that exactly the same Hamiltonian could be obtained by imposing the Coulomb gauge $\partial_i A_i =0$.

\subsubsection{Spontaneous symmetry breaking and the Higgs mechanism}
By passing to reduced phase space, the gauge group has been eliminated. There is a residual global $U(1)$ symmetry, given by
\begin{equation}
{\bar \phi} \to \ee^{\ii \alpha} {\bar \phi} \,, \quad {\bar \phi}^* \to \ee^{-\ii  \alpha} {\bar \phi}^*\,,\quad \Pi_{{\bar \phi}} \to \ee^{-\ii \alpha}\Pi_{{\bar \phi}} \,,\quad \Pi_{{\bar \phi}^*} \to \ee^{\ii \alpha} \Pi_{{\bar \phi}^*}\,, 
\label{h6}
\end{equation}
where $\alpha$ is constant. There is a degenerate ground state, given by ${\bar \phi} =  v \ee^{\ii \theta}/\sqrt{2}$, ${\bar \phi}^* =  v \ee^{- \ii \theta}/\sqrt{2}$, where $\theta$ is constant, and $\Pi_{{\bar \phi}}=\Pi_{{\bar \phi}^*}=A^T_i=\Pi_{A^T_i}=0$. The residual symmetry is regarded as a physical symmetry and can be spontaneously broken. 

In order to consider perturbations around the vacuum ${\bar \phi} =  v/\sqrt{2}$, it is convenient to perform the canonical transformation
\begin{align}
{\bar \phi} &= \frac{1}{\sqrt{2}} (v + \eta + \ii \xi) \,, & \Pi_{{\bar \phi}} &= \frac{1}{\sqrt{2}} (\Pi_{\eta}  - \ii \Pi_{\xi} )\,, \nonumber\\
{\bar \phi}^* &= \frac{1}{\sqrt{2}} (v + \eta - \ii \xi) \,, & \Pi_{{\bar \phi}^*} &= \frac{1}{\sqrt{2}} (\Pi_{\eta}  + \ii \Pi_{\xi} )\,,
\label{h7}
\end{align}
where $\eta,\Pi_{\eta},\xi,\Pi_{\xi}$ are real. The perturbations are described by small fields $\eta$ and $\xi$. Substitution in the Hamiltonian yields, up to quadratic terms:
\begin{multline}
H = \int d^3 x \bigg(\frac{1}{2} \Pi^2_{\eta}  + \frac{1}{2} \partial_i \eta \partial_i \eta + \mu^2 \eta^2 + \frac{1}{2} \Pi^2_{\xi}  - \frac{e^2v^2}{2}\Pi_{\xi}\frac{1}{\nabla^2} \Pi_{\xi} + \frac{1}{2} \partial_i \xi \partial_i \xi   \\
+  \frac{1}{2} \Pi_{A^T_i}\Pi_{A^T_i} + \frac{e^2v^2}{2} A^T_iA^T_i   - \frac{1}{2} A^T_i \nabla^2 A^T_i  \bigg)\,. 
\label{h8}
\end{multline}
This can be brought in a more familiar form by considering $\partial_i \xi$ as the longitudinal component of the vector potential, that is,
\begin{equation}
A^L_i = - \frac{1}{ev}\partial_i \xi , \quad \Pi_{A^L_i} = ev \frac{\partial_i}{\nabla^2}\Pi_{\xi}\,.
\label{h9}
\end{equation}
This transformation is not canonical; the Poisson bracket of $A^L_i$ and $\Pi_{A^L_i}$ is given by the longitudinal delta function. The latter variables allow us to introduce the variables 
\begin{equation}
A_i =A^T_i +A^L_i , \quad \Pi_{A_i} = \Pi_{A^T_i} + \Pi_{A^L_i}\,,
\label{h10}
\end{equation}
which form canonical pairs again. In terms of these variables the Hamiltonian reads
\begin{multline}
H = \int d^3 x \bigg(\frac{1}{2} \Pi^2_{\eta}  + \frac{1}{2} \partial_i \eta \partial_i \eta + \mu^2 \eta^2 \\
+ \frac{1}{2} \Pi_{A_i}\Pi_{A_i} + \frac{1}{2e^2v^2} \left(\partial_i \Pi_{A_i}\right)^2 + \frac{e^2v^2}{2} A_iA_i + \frac{1}{4}F_{ij}F_{ij}  \bigg)\,. 
\label{h11}
\end{multline}
The first line is recognized as the Hamiltonian of a scalar field with mass $\sqrt{2}\mu$ and the second line as that of a spin-1 field with mass $ev$. So the standard results are obtained.

\section{Acknowledgments}
It is a pleasure to thank Gordon Belot, Richard Healey, Christian Maes, Rafael Sorkin, Koen Struyve, Walter Troost and Andr\'e Verbeure for stimulating discussions and helpful comments. Support of the FWO-Flanders is acknowledged.

\appendix
\section{Proofs for section \ref{etymt}}
Denote ${\bar {\mathcal G}}^\infty$ the group of smooth maps from ${\mathbb R}^3$ to $G$ that go to the identity at spatial infinity and ${\bar {\mathcal G}}^\infty_0$ the component of ${\bar {\mathcal G}}^\infty$ continuously connected to the identity. We have that if $g \in {\mathcal G}^\infty$, with $g|_{t=t_0} \in {\bar {\mathcal G}}^\infty_0$, for some time $t_0$, then $g \in {\mathcal G}^\infty_0$. Namely, if $g|_{t=t_0} \in {\bar {\mathcal G}}^\infty_0$, then the function $g_{t_0} \in {\mathcal G}^\infty$, defined by $g_{t_0}(x)=g({\bf x},t_0)$ is an element of ${\mathcal G}^\infty_0$. Since $g$ is in the same component of ${\mathcal G}^\infty$ as $g_{t_0}$ (just consider the homotopy $h(x,\tau): {\mathbb R}^4 \times [0,1] \to G$ with $h(x,\tau)=g({\bf x},\tau(t-t_0) + t_0)$, where clearly $h(x,\tau) \in {\mathcal G}^\infty$ for each value of $\tau$), we have that $g \in {\mathcal G}^\infty_0$. Hence $\pi_0({\mathcal G}^\infty) \cong \pi_0({\bar {\mathcal G}}^\infty)$.

Let us now show that ${\mathcal G}^g = {\mathcal G}^\infty_0$. Consider an element $g \in {\mathcal G}^b$ that preserves some initial data, say at $t=t_0$. So $g$ is a generator of ${\mathcal G}^g$. From $\lim_{r\to\infty} g|_{t=t_0}=1$ it follows that $\lim_{r\to\infty} g=1$ at all times, because of the assumed boundary conditions. As such, $g \in {\mathcal G}^\infty$. Further, since $g|_{t=t_0}=1$, it follows that $g|_{t=t_0}=1 \in {\bar {\mathcal G}}^\infty_0$ and the result in the previous paragraph implies that $g \in {\mathcal G}^\infty_0$. Hence ${\mathcal G}^g \subset {\mathcal G}^\infty_0$. Assume now $g \in {\mathcal G}^\infty_0$. Then $g$ is continuously connected to the identity in ${\mathcal G}^\infty$. Because $g$ is smooth, g can also be smoothly connected to the identity. That is, there exists a smooth function $h:{\mathbb R}^4 \times [0,1] \to G$, such that $h(x,0)=g$ and $h(x,1) = 1$, where $h(x,\tau) \in {\mathcal G}^\infty$ for all $\tau \in [0,1]$ (cf.\ proposition 10.20 in \cite{lee03}). As such, there also exists a smooth function ${\bar h}:{\mathbb R}^4 \times {\mathbb R} \to G$, such that ${\bar h}(x,\tau) =g$, for $\tau \leqslant 0$, and ${\bar h}(x,\tau) = 1$, for $\tau \geqslant 1$, and $h(x,\tau) \in {\mathcal G}^\infty$ for all $\tau \in {\mathbb R}$. Now write $g=g_1g_2$ with $g_1(x)={\bar h}(x,t)$ and $g_2(x)={\bar h}(x,t)^{-1} g(x)$. As such $g_1$ and $g_2$ are both elements of ${\mathcal G}^g$, because they both preserve some initial data (since $g_1 = 1$ for $t \geqslant 1$ and $g_2 = 1$ for $t \leqslant 0$). Hence $g \in {\mathcal G}^g$ and ${\mathcal G}^\infty_0 \subset {\mathcal G}^g$.

\end{document}